\documentstyle[a4,12pt]{article}
\begin{document}
\hoffset 0.25in
\setcounter{page}{1}

\newcommand{\biindice}[3]%
{
\renewcommand{\arraystretch}{0.5}
\begin{array}[t]{c}
#1\\
{\scriptstyle #2}\\
{\scriptstyle #3}
\end{array}
\renewcommand{\arraystretch}{1}
}
\newcommand{\be}{\begin{equation}}
\newcommand{\ee}{\end{equation}}
\newcommand{\bea}{\begin{eqnarray}}
\newcommand{\eea}{\end{eqnarray}}
\newcommand{\nn}{\nonumber}
\newcommand{\muh}{\hat\mu}
\newcommand{\dlr}{\stackrel{\leftrightarrow}{D} _\mu}
\newcommand{\vnew}{$V^{\rm{NEW}}$}
\newcommand{\vecp}{$\vec p$}
\newcommand{\dof}{{\rm d.o.f.}}
\newcommand{\prd}{Phys.Rev. \underline}
\newcommand{\pl}{Phys.Lett. \underline}
\newcommand{\prl}{Phys.Rev.Lett. \underline}
\newcommand{\np}{Nucl.Phys. \underline}
\newcommand{\vvp}{v_B\cdot v_D}
\newcommand{\dl}{\stackrel{\leftarrow}{D}}
\newcommand{\dr}{\stackrel{\rightarrow}{D}}
\newcommand{\mev}{{\rm MeV}}
\newcommand{\gev}{{\rm GeV}}
\newcommand{\calp}{{\cal P}}
\pagestyle{empty}
\begin{flushright}
IPNO/TH 95-05  \\
LPTHE-ORSAY 95-06  \\
January 1995
\end{flushright}
\vskip 1.5cm
\centerline{\LARGE{\bf{Painlev\'e Analysis and Integrability Properties}}}
\vskip 0.1cm
\centerline{\LARGE{\bf{of a 2+1 Nonrelativistic Field Theory}}}
\vskip 1.5cm
\centerline{\bf{M. Knecht$^{\rm a}$, R. Pasquier$^{\rm a}$, J. Y.
Pasquier$^{\rm b}$}}
\vskip 0.5cm
\centerline{$^a$ IPN, Division de Physique Th\'eorique, F-91406 Orsay Cedex,
France,\footnote
{Unit\'e de Recherche des Universit\'es Paris XI et Paris VI associ\'ee au
CNRS.}}
\centerline{$^b$ LPTHE, F-91405 Orsay Cedex, France,\footnote {Unit\'e de
Recherche de
l'Universit\'e Paris XI associ\'ee au CNRS.}}
\vskip 2.5cm
\date{1994}
%\maketitle

 %\newpage
\begin{abstract}
 A model for planar  phenomena introduced by Jackiw and Pi and described by a
Lagrangian
including a Chern-Simons term is considered. The associated equations of
motion, among
which a 2+1 gauged nonlinear Schr\"odinger  equation, are rewritten into a
  gauge independent form involving the modulus of
the matter field.  Application of a Painlev\'e  analysis, as adapted to partial
differential
equations by Weiss, Tabor and Carnevale,  shows up  resonance values that are
all
integer. However, compatibility conditions need be  considered which cannot be
satisfied
consistently in general. Such a result  suggests that the examined equations
are not integrable, but
 provides tools  for the investigation of  the integrability  of
different reductions. This in particular puts forward the familiar integrable
Liouville and 1+1
nonlinear Schr\"odinger equations.
 \end{abstract}
\vskip 1.5cm
\begin{flushleft}
\end{flushleft}
\newpage
\pagestyle{plain}
\section{Introduction}
\label{sec:intro}
Field theories  involving  Chern-Simons terms have been
 thought to play a role in the description of planar
  phenomena, among which the fractional quantum Hall effect and
the high-$T_c$ superconductivity [1].
\par Here we are concerned with a model introduced by Jackiw and Pi [2]
which describes a self-interacting non-relativistic matter field $\psi\equiv
\psi(x_\nu),\ \nu=0,1,2$, coupled to an abelian gauge field
 $A_\mu\equiv A_\mu(x_\nu),\ \mu,\ \nu=0,1,2$. The
classical equation of motion for $\psi$ is a gauged 2+1 nonlinear Schr\"odinger
equation
$$iD_0\psi+\frac{1}{2m}{\bf D}^{2}\psi + g(\psi^*\psi)\psi=0\eqno{(1a)}$$
with
$$D_0\equiv\partial_0-iA_0, \  \ {\bf D}\equiv\nabla-i{\bf A}.$$
In the corresponding Lagrangian, the standard Maxwell kinetic term for the
gauge field is replaced by a Chern-Simons term [3]
$$-\frac{\kappa}{2}\epsilon^{\mu\nu\rho}A_\mu\partial_\nu A_\rho ,$$
so that Eq.(1a) is supplemented by ($i,\ j=1,2$)
$$\psi^*\psi-\kappa\epsilon_{ij}\partial_iA_j=0,\eqno{(1b)}$$
$$J_i-\kappa\epsilon_{ij}(\partial_jA_0-\partial_0A_j)=0,\eqno{(1c)}$$
with
$$J_i\equiv\frac{i}{2m}[(D_i\psi)^*\psi-\psi^*(D_i\psi)]$$
 (the equations dealt with in [2] are recovered by setting
$A_0\rightarrow-A_0,\
\kappa\rightarrow-\kappa$). The fields $A_\mu$  are indeed purely
auxiliary: upon fixing the gauge, they can be expressed in terms of
$\psi^*\psi$ and of the
current $\bf{J}$. Eq.(1a) then becomes a nonlinear integro-differential
equation
for $\psi$. Another peculiarity of the model lies in the fact that exact
 solutions associated with {\em static} self dual (or anti self dual)
configurations can be
constructed for $mg |\kappa| = 1 $ [2].
 \par However, important questions  still remain unanswered concerning the
integrability
properties of this model. In this respect,  a Painlev\'e analysis in a form
adapted to partial differential equations  by Weiss, Tabor and Carnevale (WTC)
[4]  may be a source of
information.
\par  Such an analysis is performed below in Section 2. It gives  evidence that
Eqs.(1)
 are {\em not} integrable, although they naturally admit integrable
reductions among which the familiar Liouville and 1+1 nonlinear Schr\"odinger
equations.

\section{The Painlev\'e-WTC analysis}
\label{sec:Pain}
To simplify matters, it is first worthwhile to  introduce
dimensionless fields and coordinates defined by
$$\psi=|\kappa|^{\frac{3}{2}}re^{i\chi},\
A_0=\frac{\kappa^2}{2m}w+\partial_0\chi,
\ A_1=\kappa u+\partial_1\chi,\ A_2=\kappa v+\partial_2\chi,$$
and by
$$x_0=\frac{2m}{\kappa|\kappa|}t,\ \ x_1=\frac{x}{|\kappa|},\ \
x_2=\frac{y}{|\kappa|}
.$$
The new (real) fields $u\equiv u(x,y,t), \ v\equiv v(x,y,t),\ w\equiv
w(x,y,t),\
r\equiv r(x,y,t) $ satisfy equations  readily deduced from (1). For further
convenience, these are
 ordered as
$$u_y-v_x  =  - r^2, \eqno{(2a)}$$
$$u_t-w_x  =  -2 v r^2,  \eqno{(2b)}$$
$$v_t-w_y  =  2 u r^2,J\eqno{(2c)}$$
$$r_{xx}+r_{yy}  =  2 C r^3-r(w-u^2-v^2), \eqno{(2d)}$$
$$(r^2)_t  =  2(u r^2)_x + 2(v r^2)_y,\eqno{(2e)}$$
with $C\equiv 1/B=-mg |\kappa|$ and, as usual, $u_y=\partial_yu, \
v_x=\partial_xv,$ etc.
\par Note that in the static limit
(no time dependence), with moreover $C^2\equiv 1/B^2=1$, Eqs.(2) are associated
with the
configurations considered in [2]. In this
case, they admit the particular solutions $u = -B (r_y/r),\ v =B (r_x/r),\ w =
B r^2$, with $r$
satisfying the  2+0 Liouville equation
	$${\cal L} (r) \equiv (\frac{r_x}{r})_x + (\frac{r_y}{r})_y - C r^2
= 0\eqno{(3)}$$
(indeed the physical, zero energy, solutions are obtained for $C\equiv
1/B=-1$).
\par On the other hand, if all functions in Eqs.(2) only
depend on $x,\ y$  through the combination $\xi = x+ \gamma_c y,$ with $
\gamma_c$ a constant
that  may be zero, then we can introduce  $\theta\equiv \theta(\xi ,t)$ such
that $\theta_\xi =
(u+\gamma_c v)/(1+\gamma_c^2),\ \theta_t = w-u^2-v^2 + (u+\gamma_c v)^2
/(1+\gamma_c^2)$.
In addition,   $\Psi(\xi,t) \equiv r(\xi,t) \exp{i\theta(\xi,t)}$  can be shown
to satisfy the
1+1 nonlinear Schr\"odinger equation (NLSE)
$$- i\Psi_t + (1+\gamma_c^2)\Psi_{\xi\xi} = 2C |\Psi|^2\Psi .\eqno{(4)}$$
\par Let us also emphasize that the phase $\chi$ of $\psi$, which  is closely
related to  the
gauge invariance properties of (1), no  longer appears in the {\em gauge
independent} system (2).
Furthermore, (2e) is nothing else  than the compatibility condition of (2a-c)
and amounts to the
continuity equation $\partial_0(\psi^*\psi)+\nabla.{\bf J}=0$ associated with
(1).
 \par Therefore, any Painlev\'e WTC analysis
of (1) may be restricted to (2a-d). At the same time, it is interesting to
examine whether this
allows the afore mentioned  reductions (3), (4) to be put forward.
\par As in [4], we perform this analysis by  looking for solutions $u,\ v,\ w,\
r$
 whose behaviour about any movable non characteristic singular manifold
$\phi\equiv
\phi(x,y,t) = 0$ is defined  by generalized Laurent series
$$u=\sum_{j=0}^\infty u_j\phi^{j-p_1},\ v=\sum_{j=0}^\infty v_j\phi^{j-p_2},
\ w=\sum_{j=0}^\infty w_j\phi^{j-p_3},
\ r=\sum_{j=0}^\infty r_j\phi^{j-p_4},\eqno{(5a)}$$
with {\em a priori} unknown leading powers  $p_1,\ p_2,\ p_3,\ p_4$. As usual,
we
 suppose that the  coefficients $u_j\equiv
u_j(x,y,t),\ v_j\equiv v_j(x,y,t),\ w_j\equiv w_j(x,y,t),\ r_j\equiv
r_j(x,y,t)$ only
 depend on the derivatives of $\phi$, with e.g. $ \phi_x\not=0$ when $\phi =0$.
Moreover, we
assume that the relevant relations satisfied by these coefficients are merely
obtained  by  inserting
(5a) into (2) and by balancing the contributions at each order of $\phi$ in the
resulting equations
$$\sum_{k=0}^\infty E_k^{(i)} \phi^{k-q_i}=0,\eqno{(5b)}$$
 $i=1,2,3,4$ for (2a,b,c,d).
\par Such operations, when applied to the dominant terms
corresponding to $k=0$, naturally put forward solutions that are {\em a priori}
singular
about $\phi=0$, with
$$	p_1 = p_2 = p_4 =1 ,     \ \ p_3 = 2  $$
 in (5a), and $ q_1=2, \ q_2=q_3=q_4=3$ in (5b). At the same time, they give
$$u_0=B\phi_y=B\gamma
\phi_x,\ v_0=-B\phi_x,\
w_0=B^2(\phi_x^2+\phi_y^2)=B^2(1+\gamma^2)\phi_x^2\eqno{(6a,b,c)}$$
and
$$r_0^2=B(\phi_x^2+\phi_y^2)=B(1+\gamma^2)\phi_x^2.\eqno{(6d)}$$
 %  invariant
In these expressions, $\gamma$ stands for the quantity
$$\gamma = \frac{\phi_y}{\phi_x}\eqno{(7)}$$
which is invariant under the   M\"obius group of homographic
transformations $\phi\rightarrow \frac{\alpha \phi +\beta}{\lambda \phi+\mu},$
depending on
constants $\alpha,\ \beta,\ \lambda,\ \mu$  such that
$\alpha\mu-\beta\lambda=1$ (see
e.g. [5]  for the use in other contexts of expansions that systematically put
forward the invariance
properties under this group).
\par On the other hand, for $k \geq 1$,  we obtain the relations
$$(k-1)\phi_y
u_k-(k-1)\phi_xv_k+2r_0r_k = S_k^{(1)}, \eqno{(8a)}$$  $$\ 2r_0^2
v_k-(k-2)\phi_x
w_k+4v_0r_0r_k = S_k^{(2)}, \eqno{(8b)}$$  $$-2r_0^2 u_k-(k-2)\phi_y
w_k-4u_0r_0r_k =
S_k^{(3)},\eqno{(8c)}$$  $$-2r_0
u_0u_k-2r_0v_0v_k+r_0w_k+(k^2-3k-4)(\phi_x^2+\phi_y^2)r_k =
S_k^{(4)} ,\eqno{(8d)}$$
whose second members $S_k^{(i)},\  i=1,2,3,4$, only involve  functions  $u_j,\
v_j,\
w_j,\ r_j$ with $j< k$.    More precisely
$$  S_k^{(1)}=-	u_{k-1,y}+v_{k-1,x}	-\biindice{\sum}{j+j'=k}{j,j'\neq
k}r_jr_{j'}	,\eqno{(9a)}$$
$$ S_k^{(2)} =-  u_{k-2,t}-(k-2)\phi_t
u_{k-1}+w_{k-1,x}-2\biindice{\sum}{j+j'+j"=k}{j,j',j"\neq k}
v_jr_{j'}r_{j"}		 ,\eqno{(9b)}	$$
 $$ S_k^{(3)} =-v_{k-2,t}-(k-2)\phi_t
v_{k-1}+w_{k-1,y}+2\biindice{\sum}{j+j'+j"=k}{j,j',j"\neq k}
u_jr_{j'}r_{j"}		,\eqno{(9c)}$$
 $$ S_k^{(4)} =	-(k-2)(\phi_{xx}+\phi_{yy})r_{k-1}-2(k-2)(\phi_x
r_{k-1,x}+\phi_y r_{k-1,y})$$
$$\ \ -r_{k-2,xx}	-r_{k-2,yy}
+2C\biindice{\sum}{j+j'+j"=k}{j,j',j"\neq k}r_jr_{j'}r_{j"}	$$
$$\ -\biindice{\sum}{j+j'=k}{j,j'\neq k}r_jw_{j'}
+\biindice{\sum}{j+j'+j"=k}{j,j',j"\neq
k}r_j(u_{j'}u_{j"}+ v_{j'}v_{j"}).\eqno{(9d)}	$$
\par In fact, Eqs.(8) define an algebraic linear system for   $u_k
,\ v_k,\ w_k,\ r_k$.  Owing to (6),  the associated determinant can
  be calculated and reduced to
$$Det(k)=-2B(\phi_x^2+\phi_y^2)^3(k+1)(k-1)(k-2)(k-4).\eqno{(10a)}$$
In this calculation and the following ones, both branches of $r_0$ in (6d) may
indeed be treated
simultaneously, which amounts to  working with $r^2$ instead of $r$.
\par Note that $Det(k)$ vanishes at values $k_r$ of $k$ that are all integer,
namely
$$k_r=-1,1,2,4.\eqno{(10b)}$$
Among these ``resonance values'', $k_r=-1$ as usual reflects the
arbitrariness of $\phi$. For the other values of $k_r$, we have to check
whether the
equations are compatible {\em per se} or only for restricted $\phi$
satisfying some ``consistency condition'' [4]. In either case, since all roots
of
(10a) are simple, one of the functions $u_{k_r},\ v_{k_r},\ w_{k_r},\  r_{k_r}$
remains
undetermined and correspondingly appears as  arbitrary in the expansions (5a).
\par Let us add that,  in connection with (3), it is also useful to write
	$$u = - B b + U ,\ v =B a + V ,\ w = B \rho + W,\eqno{(11a)}$$
with
	$$a=\frac{r_x}{r},\  b= \frac{r_y}{r},\ \rho=r^2.\eqno{(11b)}$$
Correspondingly, we  have
$$	u_k = -B b_k +U_k,\ v_k = B a_k +V_k, \ w_k = B\rho_k
+W_k,\eqno{(11c)}$$
where   $a_k,\ b_k,\ \rho_k$ and $U_k,\ V_k,\ W_k$ are the coefficients
involved in $\phi$
expansions like (5) for  $a,\ b,\ \rho$ and $U,\ V,\ W$. More precisely,
$a_k,\ b_k,\ \rho_k$
can be determined recursively from the $r_l,\ l\leq k$, by identification of
powers of $\phi$ in
(11b). The recurrence relations for  $U_k,\ V_k,\ W_k$   follow  by similar
operations in the
equations  for  $U,\ V,\ W$  obtained by inserting (11a) into (2).
Alternatively, these relations
may be deduced by substituting (11c) into (8), (9).
  \par In fact,  a splitting such as (11a) advantageously puts forward - and
indeed yields
for $U=V=W=0$ - the form (11b) of the solutions associated with the $C^2\equiv
1/B^2=1$ static
configurations examined in [2]. Correspondingly, a part $r_L$ that satisfies
the Liouville equation
(3) could have been separated off $r$, e.g. by writing $r=r_L+R$, with  $ R=0$
for the solutions
 in  [2]. However, for the present purpose, we  only need to use  $U,\ V,\ W$
so as to cast
  the results in a sufficiently simple and  structured form. In this respect,
note that
owing to (6) and to the fact that
 $$a_0 = -\phi_x  ,\ b_0 = -\phi_y  ,\ \rho_0 = r_0^2 =B
(\phi_x^2+\phi_y^2),\eqno{(12)}$$
we simply have  $ U_0=V_0=W_0= 0$ for $k=0$.
\par Let us now  examine the cases $k=1,\ 2,\ 3,\ 4$ successively.
\subsection{The case $k = 1$ (resonance)}
Owing to (9), the system (8) for $k=1$ reduces  to
$$2r_0r_1 = -u_{0y}+v_{0x}, \eqno{(13a)}$$
$$\ 2r_0^2v_1+\phi_x w_1+4v_0r_0r_1 = \phi_tu_0+w_{0x}, \eqno{(13b)}$$
$$-2r_0^2u_1+\phi_y w_1-4u_0r_0r_1 = \phi_tv_0+w_{0y},\eqno{(13c)}$$
$$-2r_0u_0u_1-2r_0v_0v_1+r_0w_1 -6(\phi_x^2+\phi_y^2)r_1 =
(\phi_{xx}+\phi_{yy})r_0
 +2(\phi_x r_{0x}+ \phi_y r_{0y}),\eqno{(13d)}$$
with $u_0,\ v_0,\ w_0,\ r_0$ given by (6).
\par In fact, Eqs.(13b-d)   have a vanishing
determinant (therefore  the whole system (13) too) and  turn out
to be compatible only if
$$	(B^2-1) (\phi_{xx}\phi_y^2 - 2\phi_x\phi_y\phi_{xy}+ \phi_{yy}\phi_x^2)=0.
\eqno{(14a)}$$
Up to factors which are generally  non vanishing,  this condition is
equivalent to (cf. (7))
$$	C_1\equiv   (B^2-1) (\gamma_y-\gamma \gamma_x) = 0.\eqno{(14b)}$$
Hence, the compatibility is ensured  if $B^2 \equiv 1/C^2= 1$ or/and if
$$	\phi_{xx}\phi_y^2 - 2\phi_x\phi_y\phi_{xy}+ \phi_{yy}\phi_x^2=0, \  i.e. ,\
\gamma_y-\gamma \gamma_x=0.\eqno{(15a,b)}$$
Note that (15a) is just the Bateman equation in two variables [6] whose
determinantal form is
$$det
\left(
\begin{array}{ccc}
0            &  \phi_x           &  \phi_y    \\
\phi_x  &  \phi_{xx}     &   \phi_{xy}  \\
\phi_y  &  \phi_{yx}     &   \phi_{yy}
\end{array}
\right)
=0.$$
\par	At the end, if the ``consistency condition'' (14) - hereafter referred to
as (C1) - is satisfied,
then Eqs.(13b-d) determine two among the  unknowns $u_1 ,\ v_1 ,\ w_1$ in terms
of the
third one which remains arbitrary. For further convenience, the result is here
written as (cf. (11c))
$$	u_1 = -B b_1 +U_1,\ v_1 = B a_1 +V_1,\ w_1 = B\rho_1 +W_1, \eqno{(16a)} $$
with
$$a_1 = \phi_x \frac{r_1 }{r_0} +\frac{ r_{0 x}}{r_0} ,\ \ b_1 =
 \phi_y\frac{r_1 }{r_0} +\frac{ r_{0 y}}{r_0} ,
\ \rho_1 = 2 r_0 r_1 , \eqno{(16b)} $$
$$U_1 =\frac{\Gamma -\gamma {\cal  H}} {2} ,
\ V_1 = \frac{ {\cal H} }{2} ,
\ W_1= B\phi_x[\gamma \Gamma  - {\cal  H}(1+\gamma^2)],\eqno{(16c)} $$
and, from (13a), (6a,b),
$$	\rho_1 = 2 r_0 r_1 = - B(\phi_{xx}+\phi_{yy}).\eqno{(16d)}$$
In these expressions, $r_0$ has still to be replaced owing to (6d) while
${\cal H}\equiv {\cal
H}(x,y,t)$  is an arbitrary function proportional to  $V_1$. On the other hand,
$\Gamma$  is the
M\"obius invariant
$$\Gamma = \frac{\phi_t}{\phi_x}\eqno{( 17a)}$$
which, owing to  the required identity $(\phi_t)_y\equiv (\phi_y)_t$
 (compatibility of the definitions (17a) and (7)), is such that
$$\Gamma_y-\gamma \Gamma_x \equiv \gamma_t-\Gamma
\gamma_x.\eqno{( 17b)}$$
\subsection{The case $k = 2$ (resonance)}
In this case, Eqs.(8) and  (9) give
$$\phi_y u_2-\phi_x v_2+2r_0r_2 = -u_{1y}+v_{1x}-r_1^2, \eqno{( 18a)}$$
$$2r_0^2 v_2+4v_0r_0r_2 = -u_{0t}+w_{1x}-2(r_1^2v_0+2v_1r_0r_1), \eqno{(
18b)}$$
$$-2r_0^2 u_2-4u_0r_0r_2 = -v_{0t}+w_{1y}+2(r_1^2u_0+2u_1r_0r_1),\eqno{(
18c)}$$
$$-2r_0 u_0u_2-2r_0v_0v_2+r_0w_2-6(\phi_x^2+\phi_y^2)r_2 =
-(r_{0xx}+r_{0yy}) $$
$$\ \ +6C r_1^2r_0-w_1r_1+r_0(u_1^2+v_1^2)+2r_1 (u_0u_1+v_0v_1).\eqno{( 18d)}$$
In fact, Eqs.(18a-c)  have a vanishing determinant
(therefore the whole system (18) too) and turn out to be compatible
only if
$$2B(\phi_x^2+\phi_y^2)S_2^{(1)}+\phi_xS_2^{(2)}+\phi_yS_2^{(3)}=0\eqno{(19a)}$$
where the $S_2^{(i)},\ i=1,2,3,$ stand for their second members.
\par This new condition is hereafter referred to as (C2). Owing to (6), (7),
(16) and (17), and up to
global factors which are generally  non vanishing,
 it simply reduces to
$$	C_2\equiv \Gamma_y-\gamma \Gamma_x -{\cal H}
(\gamma_y-\gamma \gamma_x) =0, \eqno{(19b)}$$
where $\Gamma_y-\gamma \Gamma_x$ may be replaced by $\gamma_t-\Gamma
\gamma_x$ just as well (cf. (17b)).  In the general case, with an  arbitrary
${\cal H}$, this requires
  $\gamma_y=\gamma
\gamma_x$ (Eq.(15b)),  {\em and}
$$ \Gamma_y=\gamma \Gamma_x,\	\mbox{or equivalently,}\  \gamma_t=\Gamma
\gamma_x.\eqno{(20a,b)}$$
Note that, in terms of the derivatives of $\phi$, the latter condition also
reads as
$$	2 \phi_x\phi_t(\phi_{yt}\phi_x -\phi_{xt}\phi_y)+ \phi_t
(\phi_{xx}\phi_y^2 -  \phi_{yy}\phi_x^2)=0$$
and  amounts to the determinantal three variable equation [7]
$$det
\left(
\begin{array}{cccc}
0            &  \phi_x           &  \phi_y         &  \phi_t       \\
\phi_x  &  \phi_{xx}     &   \phi_{xy}   &   \phi_{xt}  \\
\phi_y  &  \phi_{yx}     &   \phi_{yy}  &   \phi_{yt}  \\
\phi_t  &  \phi_{tx}     &   \phi_{ty}  &   \phi_{tt}
\end{array}
\right)
=0,$$
restricted by (15a).
\par	At the end,  provided that (19) holds, Eqs.(18a-d) yield for instance
$u_2,
\ v_2,\ w_2$ in terms of $r_2$  taken arbitrary and assumed, without loss of
generality, to be
independent of   $\Gamma$ and ${\cal H}$. The result, when  written
in the form (11c), involves
$$a_2=2\frac{r_2}{r_0}\phi_x+\frac{r_{1x}}{r_0}
-\frac{r_1}{r_0}a_1,\ b_2=2\frac{r_2}{r_0}\phi_y+\frac{r_{1y}}{r_0}
-\frac{r_1}{r_0}b_1,$$
$$\rho_2 = 2 r_0r_2 + r_1^2.\eqno{(21a)}$$
Moreover, we find that
$$	U_2\phi_x = -\frac{ (\Gamma -\gamma {\cal H})_y }{ 2\gamma}   ,
\ V_2\phi_x  =  - \frac{ {\cal H}_x}{2} ,  \ W_2= (B^2-1) W_2^B + W_2^{nB},
\eqno{(21b,c,d)}$$
with
$$	W_2^B  =a_1^2+b_1^2+2a_0a_2+2b_0b_2-C\rho_2,\eqno{(21e)}$$
 $$W_2^{nB} = - 2 B(U_{1y} - V_{1x}) + U_1^2+V_1^2 -
 2 B (U_1b_1-  V_1a_1).\eqno{(21f)}$$
In all these expressions, $r_0, \ a_0 , \ b_0 , \ a_1 , \ b_1 , \ U_1 , \ V_1
,\ r_1$ have to be explicited
according to (6d), (12) and (16b-d). In fact, owing to the  factor $(B^2-1)$,
the $\Gamma$- and ${\cal
H}$-independent part $W_2^B $  needs to be evaluated for $\gamma_y=\gamma
\gamma_x$	only (the
constraint (C1) requires that every power or derivative of
$(B^2-1)(\gamma_y-\gamma \gamma_x)$
cancels). This simply yields
$$W_2^B  =- 6 C r_0r_2  - 6 C r_1^2+\frac {r_{0xx}+r_{0yy}}{r_0}.\eqno{(21g)}$$
 Similarly, owing to (C2), the combinations $(\Gamma_y-\gamma
\Gamma_x)$ and $(\gamma_t-\Gamma
\gamma_x)$ may be replaced by ${\cal H}(\gamma_y-\gamma \gamma_x)$.
\subsection{The case $k = 3$ }
 For $k=3$, the system (8) becomes
$$2\phi_y u_3-2\phi_x v_3+2r_0r_3= S_3^{(1)},  \eqno{(22a)} $$
$$\ 2r_0^2 v_3-\phi_x w_3+4v_0r_0r_3= S_3^{(2)},  \eqno{(22b)} $$
$$-2r_0^2 u_3-\phi_y w_3-4u_0r_0r_3 = S_3^{(3)}, \eqno{(22c)} $$
$$-2r_0 u_0u_3-2r_0v_0v_3+r_0w_3-4(\phi_x^2+\phi_y^2)r_3= S_3^{(4)} ,
\eqno{(22d)} $$
 and has a non vanishing determinant, cf.(10). Hence, this system admits a
unique solution
which may be readily expressed in
terms of the $S_3^{(i)}$ and thus, owing to (9),  in terms  of the  $u_k,\
v_k,\ w_k,\
r_k,\ k\leq 2,$ determined previously. In correspondance with (11c), (16) and
(21), splittings
 such as
$$	u_3 = -B b_3 +U_3,\ v_3 = B a_3 +V_3, \ w_3 = B\rho_3
+W_3,\eqno{(23a)}$$
may also be considered, together with structures  in the $U_3,\ V_3,\ W_3$
that have well defined behaviours at relevant limits such as $B^2= 1$ or/and
$\Gamma = 0$ (static
limit) or/and ${\cal H}=0$. In particular,  it is useful to write
$$	U_3 = (B^2-1) U_3^B+ U_3^{nB},    \ V_3 = (B^2-1) V_3^B+ V_3^{nB},
 \ W_3 =(B^2-1)W_3^B+ W_3^{nB},\eqno{(23b)} $$
where all contributions independent of  $\Gamma$ and ${\cal H}$ are gathered in
the
 $U_3^B,\ V_3^B,\ W_3^B$. Note that these $B$-indexed parts, already put
forward in Eq.(21d) and
such that $U_k^B= V_k^B=0$  for  $k\leq 2$, $W_k^B= 0$  for $k\leq 1$,
 are at the same time multiplied by an overall factor $(B^2-1)$. Owing to their
definition,
they are found again   in an analysis of static forms of Eqs.(2) which for
the $\ U,\ V,\ W$ read as (cf. (11a,b))
$$U_y - V_x = B  {\cal L}(r),
\ W_x = 2  V r^2,
\ W_y =- 2  U r^2,    $$
$${\cal L}(r)    =  (B^2-1) (a^2+b^2- C r^2) -W+U^2+V^2+2B(Va-Ub),\eqno{(24)}$$
with ${\cal L}(r) $  given by (3) or  ${\cal L}(r) \equiv a_x+b_y-C\rho$.
\par In any case, the
expressions of  $a_3,\ b_3,\ \rho_3=2(r_0r_3+r_1r_2),\ U_3,\ V_3,\ W_3,\ r_3$
 become quite lengthy - and therefore are not explicited here - when everything
is replaced in terms of the arbitraries, besides basic
$x$-derivatives of $\phi$ and M\"obius invariants. As Eq.(21g), they involve
  the Schwartzian
$S=\frac{\phi_{xxx}}{\phi_x}-\frac{3}{2}\frac{\phi_{xx}^2}{\phi_x^2}$  whose
derivatives $S_y,\ S_t$ can be expressed in terms of $S_x$  thanks to  (7) and
(17a).
\subsection{The case $k = 4$ (resonance)}
The corresponding Eqs.(8) become
$$3\phi_y u_4-3\phi_x v_4+2r_0r_4 = S_4^{(1)},  \eqno{(25a)} $$
$$2r_0^2 v_4-2\phi_x w_4+4v_0r_0r_4 = S_4^{(2)},  \eqno{(25b)} $$
$$-2r_0^2 u_4-2\phi_y w_4-4u_0r_0r_4 = S_4^{(3)}, \eqno{(25c)} $$
$$-2r_0 u_0u_4-2r_0 v_0v_4+r_0w_4= S_4^{(4)} , \eqno{(25d)} $$
where the $ S_4^{(j)}, \ j=1,2,3,4,$ can be expressed thanks to (9) in terms of
the  $u_k,\ v_k,\ w_k,\ r_k,\ k\leq 3$,  obtained previously, or  directly in
terms of the $
S_k^{(j)},\ k\leq 3$. In fact, Eqs.(25) have a vanishing determinant (cf.(10))
and are compatible
only if
$$	C_4\equiv  2r_0^2 S_4^{(1)}
+ \phi_x S_4^{(2)}+\phi_x \gamma S_4^{(3)}+2Cr_0 S_4^{(4)}=0.\eqno{(26)}$$
The analysis of this  last condition - referred to as (C4) below -  {\em a
priori}
appears quite involved.  Nevertheless, large simplifications occur which we
have also checked with
the help of computer programs (MATHEMATICA, REDUCE) allowing symbolic
manipulations. In this
respect, it is advantageous to take into account the structures put forward in
(11c), (21d) and (23),
combined with the preceding constraints (C1), (C2). At the end we obtain
	$$C_4 \equiv (B^2-1) C_4^B + C_4^{nB}=0,\eqno{(27a)}$$
with
$$	C_4^B \equiv  \frac{2}{3(1+\gamma^2)} (\delta _{\eta})^2
W_2^B,\eqno{(27b)}$$
$$	C_4^{nB} \equiv  -  \delta _{\eta}\delta _T {\cal H},\	\mbox{or, owing to
(C2),}\ C_4^{nB} \equiv
 -  \delta _T\delta _{\eta} {\cal H}+(\delta _{\eta} {\cal H})^2,\eqno{(27c)}$$
$$\delta_{\eta}  \equiv  \partial_y -\gamma \partial_x,
\ \delta_T\equiv \partial_t- \Gamma \partial_x  -{\cal H}\delta_{\eta} .
\eqno{(28)}$$
 Again, the $\Gamma$- and ${\cal H}$-independent part $ C_4^B$, with
$W_2^B$ given by (21e), may be evaluated for $\gamma_y=\gamma \gamma_x$ (in
such
conditions  $(B^2-1)(\delta _{\eta})^2  =(B^2-1)(\partial_{yy}-2\gamma
\partial_{xy} + \gamma^2
\partial_{xx})$).
\par For further discussions, let us add  that
\par (i)  alternatively, $C_4$ may be  split into
$$C_4 \equiv  C_4^S+C_4^T,\eqno{(29a)}$$
 with
$$ C_4^S \equiv  (B^2-1)C_4^B+{\cal H}(\delta _{\eta})^2 {\cal H}+(\delta
_{\eta} {\cal H})^2,
\ C_4^T \equiv  -(\partial_t-\Gamma \partial_x)
\delta _{\eta}{\cal H}.\eqno{(29b,c)}$$
Obviously, $ C_4^T $ cancels in the static limit ($\partial_t\rightarrow 0,\
\Gamma\rightarrow 0$)
and for ${\cal H}\rightarrow 0$, while in such limits,  $C_4^S$ reduces to
$(B^2-1)C_4^B$  (cf. the
remarks in Section 2.3 concerning the B indexed parts).
\par (ii) the first two constraints (C1), (C2) (cf. (14b), (19b)) may also
 be  simply rewritten in terms of the just introduced operators $\delta_{\eta}
,\ \delta_T$ as
$$C_1\equiv (B^2-1) \delta_{\eta}\gamma = 0,\eqno{(30a)}$$
and (cf.(17b))
$$C_2\equiv	\delta_{\eta}\Gamma-{\cal H}\delta _{\eta}\gamma=\delta_T \gamma =
0.\eqno{(30b)}$$
\section{Discussion and conclusion}
 It is clear on expressions (27), (29) that $C_4$  only vanishes for specific
${\cal H}$ and
$W_2^B$ - i.e. specific ${\cal H}$ and $r_2$ owing to (21e,g) - which
contradicts the assumptions
made in steps $k=1, 2$ on the arbitrariness of these functions. Therefore,
although the analysis at
$k=1, 2$ might imply the existence of some ``conditional Painlev\'e property''
for Eqs.(2a-d) (cf.[4]),
the constraint at $k=4$ shows that generally  no such  property holds at all.
According to the point
of view developped in [4], this suggests that  Eqs.(1), (2) are {\em not}
integrable.
\par These results
agree with  others  that appeared during the completion of the present work.
Namely,
 in [8]  two ODE reductions of (1) are shown to possess the Painlev\'e
property, whereas another one,
 associated with rotational invariance, does not (in fact, owing to the above
study, the first two ODE
may be easily identified with static and similarity reductions of the
integrable  NLSE Eq.(4)).
\par Precisely, the  analysis of Section 2 provides us with tools for
investigating which reductions of Eqs.(1), (2) might be integrable, i.e. are
such that
 all conditions (C1), (C2), (C4) are identically satisfied. Let us here merely
mention that
 \par i) In the static limit obtained by dropping all time dependences,  a 2+0
reduction of (2) is got for which   all the required conditions are generally
no more satisfied.
As a direct WTC analysis  also shows, we  have in this case $C_4\equiv C_4^S$
(cf.(29)), which does not
vanish for any   ${\cal H}$. This suggests that such a reduction, indeed
equivalent to two coupled
second order equations for $r,\ w$ as well as to Eqs.(24), is no more
integrable.
 \par ii) If we furthermore
restrict ourselves in case (i) to values $C^2\equiv 1/B^2=1$ and to
solutions built with ${\cal H}=0$, then all the conditions (C1), (C2), (C4)
become satisfied whatever
$\phi(x,y)$ is. At the same time, the selected solutions are such that  $U_k ,\
V_k ,\ W_k$
 vanish for $k \leq 2$ (cf. Eqs.(12), (16c) and (21b-f) with $\Gamma={\cal
H}=0$), and indeed for any
$k$ owing to the recursion laws. Therefore, we have $U = V = W  = 0$ for such
solutions, which,
owing to (24), are clearly associated with the integrable Liouville equation
${\cal L}(r)=0$.  In
fact, the analysis in Section 2 reduces in this case to that of the 2+0 Eq.(3)
and, as it should
be, the choice ${\cal H}=0$ implies a diminution of the arbitraries  in
comparison with the
case of the full Eqs.(2) (Eq.(3) has only two resonances at  $k=-1,2$).
\par iii) On the other hand, if  all the involved quantities - among which
$\phi,\ \gamma, \
\Gamma, \ {\cal H}$ and $ W_2^B$   - only depend  on $x, \ y$ through the
combination
$\xi = x+\gamma_c y$,  $\gamma_c$ constant, then  $\delta _{\eta}\gamma = 0$
(indeed $\gamma =
\gamma_c$) and  $\delta _{\eta}\Gamma =\delta_{\eta}{\cal H} =\delta
_{\eta}W_2^B  =  0$.
 Therefore, all the resonance conditions are  fulfilled for
any $C\equiv 1/B$  and  $\phi(\xi,t)$ (cf.(30), (27)). In this case, the
analysis of Section 2 only
involves   quantities which are related (cf.(4)) to
  the phase and modulus of complex functions $\Psi(\xi,t)$
 satisfying the integrable 1+1 NLSE.
\par Independently of the remarks  (i)(ii)(iii) above,  we may also expect to
take advantage of the
above  analysis  for obtaining solutions of the  equations. In this respect, it
has been emphasized by several authors (see e.g.[9]) that the use of truncated
series
 is often fruitful, not only in  integrable, but also in non
integrable cases.
\par This problem and those arising with the use of (C1), (C2), (C4) for
disclosing the integrability properties  of different reductions of (2) - among
which those found in
connection with group symmetry properties - are examined elsewhere [10].

\end{document}